\documentclass[conference, 10pt]{IEEEtran}
\IEEEoverridecommandlockouts

\usepackage{amsmath,amssymb,amsfonts}
\usepackage{graphicx}
\usepackage{textcomp}
\usepackage{amsmath}
\usepackage{multirow}
\usepackage{makecell}
\usepackage{hyperref}       % hyperlinks
\usepackage{url}            % simple URL typesetting
\usepackage{booktabs}       % professional-quality tables
\usepackage{amsfonts}       % blackboard math symbols
\usepackage{nicefrac}       % compact symbols for 1/2, etc.
\usepackage{microtype}      % microtypography
\usepackage{xcolor}         % colors
\usepackage{algorithm}
\usepackage{algpseudocode}
\usepackage{amsmath}
\usepackage{tikz}

\usepackage{orcidlink}
\usepackage{comment}
\usepackage{amssymb}
\usepackage{braket}
\usepackage{xcolor}
\usepackage{cleveref}
\usepackage{orcidlink}
\usepackage[font=footnotesize,skip=0pt]{caption}
\def\BibTeX{{\rm B\kern-.05em{\sc i\kern-.025em b}\kern-.08em
    T\kern-.1667em\lower.7ex\hbox{E}\kern-.125emX}}

\usepackage{enumitem}

\setlist[itemize]{leftmargin=*}
\setlist[enumerate]{leftmargin=*}

\begin{document}

\title{QGCL: Quantum-Guided Clause Learning for Cryptanalytic SAT 
\vspace{-6pt}
}

\author{\IEEEauthorblockN{Walid El Maouaki\IEEEauthorrefmark{1}\IEEEauthorrefmark{2}\orcidlink{0009-0004-2339-5401}, Alberto Marchisio\IEEEauthorrefmark{1}\IEEEauthorrefmark{3}\orcidlink{0000-0002-0689-4776},
Muhammad Shafique\IEEEauthorrefmark{1}\IEEEauthorrefmark{2}\IEEEauthorrefmark{3}\orcidlink{0000-0002-2607-8135}}

\IEEEauthorblockA{\IEEEauthorrefmark{1} \normalsize eBrain Lab, Division of Engineering, New York University Abu Dhabi, PO Box 129188, Abu Dhabi, UAE\\}
\IEEEauthorblockA{\IEEEauthorrefmark{2} \normalsize Center for Cyber Security, NYUAD Research
Institute, New York University Abu Dhabi, UAE\\}
\IEEEauthorblockA{\IEEEauthorrefmark{3} \normalsize Center for Quantum and Topological Systems, NYUAD Research
Institute, New York University Abu Dhabi, UAE\\
Emails: walid.el.maouaki@nyu.edu, alberto.marchisio@nyu.edu, muhammad.shafique@nyu.edu}
\vspace{-27pt}
}

\maketitle

\begin{abstract}

Power side-channel attacks on AES exploit data-dependent physical leakage to recover secret keys, but turning noisy leakage observations into a verified AES-128 key remains a hard combinational search problem. SAT-assisted power side-channel cryptanalysis addresses this challenge by encoding AES semantics, key constraints, plaintext/ciphertext consistency, and leakage predicates as CNF, so that candidate keys must satisfy the exact cryptographic specification. These cryptanalytic SAT formulas are large and highly structured; our largest controlled AES-oriented power-SCA instances contain up to $39{,}389$ variables and $137{,}712$ clauses, making a full-formula Grover search well beyond the scale studied here and beyond currently practical near-term implementations. We propose QGCL, a Quantum-Guided Conflict-Driven Clause Learning (CDCL) framework in which Grover search is invoked only on small subformulas extracted dynamically around CDCL conflict cores. The quantum subsolver returns candidate assignments and violation scores that bias branching heuristics, while final SAT/UNSAT decisions and key verification remain classical. We evaluate QGCL on AES-oriented cryptanalytic SAT instances derived from power side-channel CNFs with leakage-derived hint configurations, measuring conflicts, restarts, decisions, and propagations. The experiments show consistent reductions in these solver-internal statistics on harder instances, with up to an $86\%$ reduction in conflicts compared with the classical conflict-learning baseline. Parameter sweeps over the number of Grover oracle calls and the subproblem size identify a regime in which a modest quantum resource allocation captures most of the observed improvement.

\end{abstract}


\section{Introduction}

Power side-channel attacks (SCAs) on AES observe data-dependent power consumption during encryption and use those traces to infer the AES-128 secret key~\cite{mohamed2013improved, dubrova2025solving, renauld2009algebraic}. A SAT-assisted attack turns this recovery task into a Boolean search problem: AES round semantics, the key schedule, plaintext/ciphertext consistency, and power-leakage predicates are encoded as conjunctive normal form (CNF)~\cite{nelson1955simplest}, and a SAT solver searches for key assignments that satisfy the hard specification. The bottleneck is therefore not only the $2^{128}$ key space, but also the repeated projection of partial or noisy leakage information onto a large, structured AES formula~\cite{aloul2002solving,biere2021proof}. Fully quantum attacks that apply Grover search over the entire key space or the entire CNF remain beyond near-term hardware limits~\cite{grassl2016applying, jaques2020implementing, sarah2024practical}. This work studies a narrower hybrid alternative: Grover search is used locally on conflict-driven subformulas inside a classical CDCL solver, helping reduce the search burden from complex clause interactions while guiding SAT-assisted AES power side-channel cryptanalysis without changing the classical correctness guarantee. Leakage-derived confidence hints may be supplied to the solver, but they are secondary to the central problem: correctly and efficiently solving the AES power-SCA CNF.

Hybrid quantum-classical approaches have been proposed across several domains as a practical way to exploit quantum subroutines while retaining the reliability, scalability, and control of classical computation~\cite{arthur2022hybrid}. In quantum machine learning, for example, hybrid quantum neural and quanvolutional models have been studied not only for learning performance but also for adversarial robustness and circuit-level design stability~\cite{el2024robqunns,el2024advqunn,el2025designing}. Related hybrid principles also appear in quantum optimization and SAT-solving approaches, where quantum routines provide candidate solutions, energy estimates, or search acceleration, while classical components manage problem decomposition, verification, or iterative refinement. QGCL adopts the same philosophy in the cryptanalytic SAT setting.

\subsection{Motivation: Complexity Scaling}
% From a complexity perspective, ML-aided cryptanalysis shifts the challenge from raw key space size to the projection step that enforces cryptographic correctness. Repeatedly solving ($F \land H_i$), where ($H_i$) encodes the ML hints used in the (i) th iteration, dominates runtime~\cite{dubrova2024solving}. Classical CDCL solvers are highly optimized, yet they may still traverse large parts of the search tree when hints are weak or partially contradictory~\cite{leventi2021cryptominisat}.
For AES power side-channel cryptanalysis, the hard part is not only the nominal $2^{128}$ key space. The solver must also project partial and noisy leakage information onto an exact AES/SCA constraint system. We write this repeated query as $F_{\mathrm{AES/SCA}}\land H_i$, where $F_{\mathrm{AES/SCA}}$ contains the exact AES and leakage-model clauses and $H_i$ contains the current leakage-derived assumptions or soft guidance. When $H_i$ is weak, incomplete, or partially inconsistent, CDCL may traverse a large search tree before learning enough clauses to isolate the feasible region~\cite{leventi2021cryptominisat, dubrova2024solving}.

A full Grover attack on the complete CNF is attractive asymptotically, as summarized in Table~\ref{tab:classical-vs-grover-oracle}, but unrealistic in practice: the required circuit depth and qubit count scale with the entire formula. In contrast, a conflict-driven search suggests a more economical use of quantum resources: most of the decision complexity is concentrated in relatively small conflict cores. If we can identify these cores on the fly and run Grover only on the induced subformulas, using a limited number of quantum calls, then the quantum regime needs to handle only tens of qubits per call rather than thousands, while still influencing the global search trajectory.

\begin{figure}[t!]
    \centering
    \includegraphics[width=0.9\linewidth]{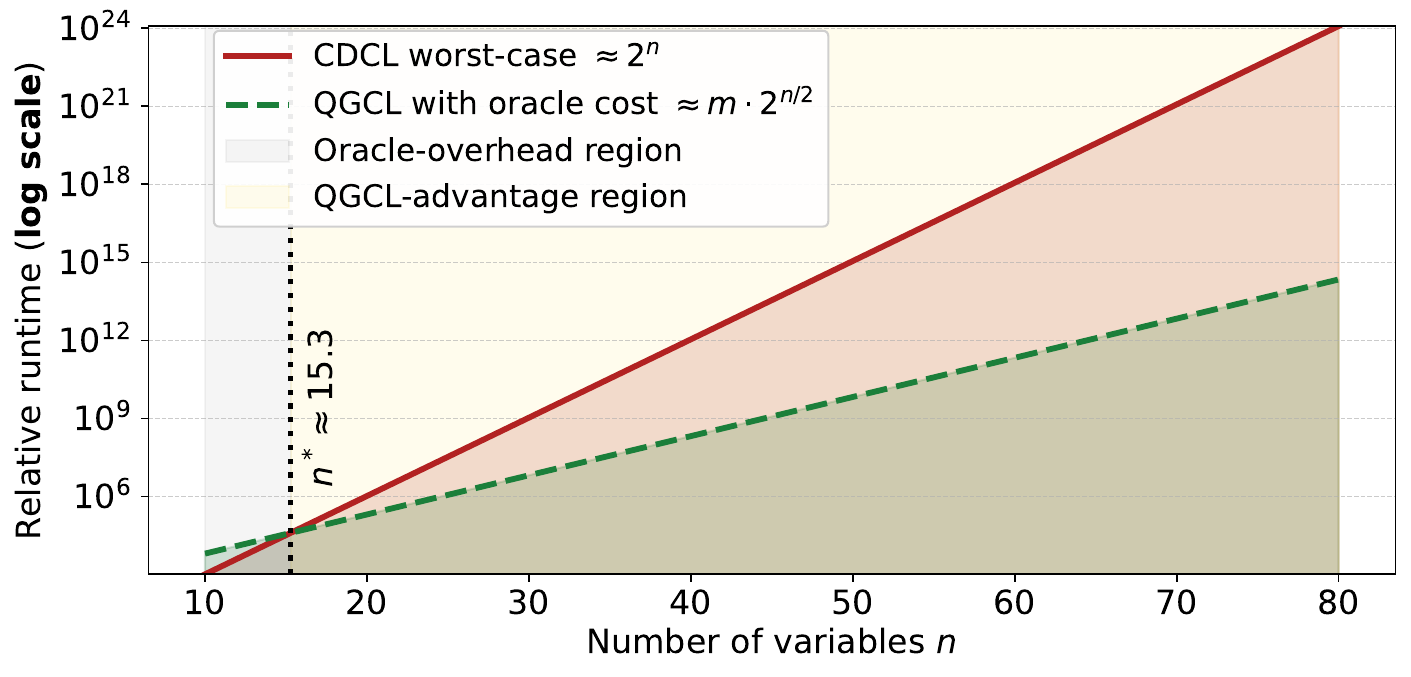}
    \caption{Asymptotic scaling motivation for applying Grover search to sufficiently large extracted CNF subformulas with $n$ Boolean variables and $m$ clauses. The red curve represents the classical worst-case term $2^n$, while the green curve represents Grover-style subformula search with clause-oracle overhead, approximately $m\cdot 2^{n/2}$. The crossover near $n^\star \approx 2\log_2 m$ marks the regime where the extracted conflict region is large enough for Grover guidance to become especially attractive. m is fixed to 200 clauses in this plot.}
    \label{Complexity}
    \vspace{-8pt}
\end{figure}

QGCL uses Grover as a targeted accelerator for the extracted-core regime. Very small extracted CNFs can be handled directly by classical SAT routines, but the same extraction mechanism can also expose conflict-rich subformulas that remain large enough for quadratic search scaling to matter. In this regime, extraction focuses the large AES/SCA formula onto the most informative variables and clauses, while Grover explores the resulting candidate space in a compact superposition. As Figure~\ref{Complexity} illustrates, comparing a classical worst-case term $2^n$ with a Grover-guided term $m\cdot 2^{n/2}$ gives a crossover near $n^\star \approx 2\log_2 m$. Therefore, for extracted subformulas beyond this crossover, QGCL can use compact quantum-state exploration to provide stronger local guidance while CDCL preserves the final correctness guarantee.

Our experiments confirm this intuition: sweeping both the number of Grover calls and the subproblem size reveals a sweet spot regime in which a small number of Grover calls on subformulas of moderate size yields a significant reduction in conflicts and restarts, without requiring large quantum circuits for the specific treated problem.

\begin{table}[t!]
\centering
\caption{Worst-case asymptotic cost of CNF-based side-channel checking with $n$ Boolean variables and $m$ clauses.}
\label{tab:classical-vs-grover-oracle}
\resizebox{0.95\columnwidth}{!}{%
\begin{tabular}{llll}
\toprule
Method & \makecell{Oracle calls /\\ iterations} & \makecell{Cost per call /\\ iteration} & Overall complexity \\
\midrule
\makecell{Grover's Algorithm\\ (unknown $K$)} 
  & $\Theta\!\big(\sqrt{2^n/K}\big)$ 
  & \makecell{Oracle: $\Theta(m)$\\ Diffuser: $\Theta(n)$} 
  & $\Theta\!\big(m\sqrt{2^n/K}\big)$ \\
\midrule
\makecell{Exhaustive\\ search~\cite{eldib2014formal}} 
  & $\Theta(2^n)$ 
  & single predicate evaluation 
  & $\Theta(2^n)$ \\
\makecell{CDCL-based SAT\\ solver~\cite{een2005minisat, leventi2021cryptominisat}}  
  & $\Theta(2^n)$ 
  & \makecell{predicate checks, BCP,\\ clause learning} 
  & $\Theta(2^n)$ \\
\makecell{SMT-based\\ verification~\cite{eldib2014smt}}  
  & $\Theta(2^n)$ 
  & \makecell{predicate checks, BCP,\\ clause learning} 
  & $\Theta(2^n)$ \\
\bottomrule
\end{tabular}%
}
\vspace{-10pt}
\end{table}

\subsection{Novel Contributions} \label{Subsec:Problem_encoding}
In this work, we introduce QGCL, a hybrid framework in which Grover search operates as a callable heuristic inside an otherwise standard CDCL solver. The quantum component is confined to small, dynamically chosen subproblems, while the classical solver retains full responsibility for the final SAT/UNSAT decision. Our main contributions are:

\begin{itemize}
    \item We propose QGCL, a conflict-local hybrid SAT framework that invokes Grover search only on bounded CNF subproblems extracted during search, rather than on the full cryptanalytic instance.
    \item We provide an engineering-level integration of quantum search with modern CDCL: conflict analysis identifies when guidance is needed, BFS-based clause expansion constructs the subproblem, and quantum outputs are translated into concrete VSIDS and polarity updates.
    \item We develop a simplified analytical model of expected hybrid runtime as a function of subproblem size, Grover-call count, and conflict reduction, clarifying the regime in which the quantum overhead is amortized by fewer classical conflicts.
    \item We give an explicit construction of Grover oracles for CNF subformulas, including clause-level ancilla management, formula-flag construction, and a BBHT-style iteration schedule suitable for bounded budgets.
    \item On AES-oriented power side-channel benchmarks with controlled problem size, including instances with up to $39{,}389$ variables and $137{,}712$ clauses, we measure both classical solver statistics (conflicts, restarts, decisions, propagations) and hybrid-control parameters (Grover budget and call count), and identify a practical sweet spot where moderate budgets capture most of the gains.
\end{itemize}

\section{Background and Related Work} 
\subsection{Problem encoding}\label{Subsec:Problem_encoding}

A Boolean satisfiability (SAT) instance consists of a propositional formula $F$ over variables $x_1, \ldots, x_n$, typically given in conjunctive normal form

\[
F(x_1,\ldots,x_n) \;=\; \bigwedge_{j=1}^{m} C_j,
\qquad
C_j \;=\; \bigvee_{i=1}^{p_j} \ell_{j,i},
\]

where each literal $\ell_{j,i}$ is either a variable $x_k$ or its negation $\neg x_k$. The SAT problem asks whether there exists an assignment
$x \in \{0,1\}^n$ such that $F(x) = 1$.

Modern high-performance solvers are based on conflict-driven clause learning~\cite{marques2009conflict}. CDCL maintains a trail of assigned literals and uses unit propagation to enforce implications of the current partial assignment. When a clause becomes falsified, the solver performs conflict analysis to derive a learned clause that prevents the same conflict from reoccurring, and backtracks to a suitable decision level. Heuristics such as Variable State Independent Decaying Sum (VSIDS), a CDCL branching heuristic based on decaying variable-activity scores~\cite{moskewicz2001chaff}, phase saving~\cite{pipatsrisawat2007lightweight}, and restart policies~\cite{huang2007effect} are crucial for practical performance, and contemporary solvers such as MiniSat~\cite{een2003extensible}, Glucose~\cite{audemard2009predicting}, and Kissat~\cite{cdfa94c389f347609408ab08c505c54a} follow this overall design.

SAT remains an NP‑complete problem~\cite{cook2023complexity}, and its worst‑case hardness underpins the difficulty of the underlying search task. In this work, SAT is the projection layer in a cryptanalytic pipeline: it enforces exact algebraic and logical constraints of cryptographic primitives given noisy information about the secret state.

\subsection{SAT Encoding for AES Power Side-Channel Attacks} \label{subsec:aes_sca_encoding}

\begin{figure*}[t!]
    \centering
    \includegraphics[width=0.8\linewidth]{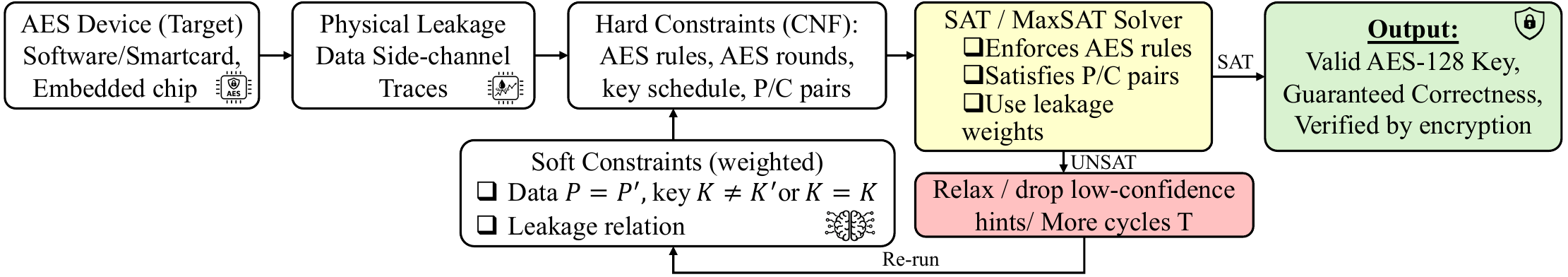}
    \caption{
SAT-assisted AES-128 power side-channel key-recovery workflow. Power traces from the target are translated into hard leakage predicates and, when confidence estimates are available, optional weighted hints. The hard CNF encodes AES round semantics, the key schedule, plaintext/ciphertext consistency, and the selected leakage relation. The SAT/MaxSAT block denotes exact SAT solving for the hard CNF, with MaxSAT-style weighting used only for optional soft hints. If the assumptions are unsatisfiable, low-confidence hints can be relaxed or additional leakage cycles $T$ can be considered. A key is accepted only after satisfying the hard constraints and passing AES encryption verification.}
    \label{fig:Problem}
\vspace{-10pt}
\end{figure*}

Figure~\ref{fig:Problem} summarizes the AES power side-channel workflow. 
Modern side-channel attacks combine recovered leakage with SAT solving. Power or EM traces are converted into simple metrics, such as bit values or Hamming weights, which encode partial state information as SAT constraints and prune the search space~\cite{renauld2009algebraic,dubrova2025solving}.

Our implementation follows an AES-128 key-recovery workflow, but the experiments use controlled 100-bit AES-oriented proxy instances to keep CNFs and Grover subproblems measurable. A full AES-128 instance would use the same workflow with 128-bit key/state width and complete AES round and key-schedule constraints. The generator fixes the plaintext, creates two symbolic key/state executions, optionally fixes known seed-key bits, and constrains the two keys to differ. In the reported benchmarks, the plaintext is one and the leakage model is Hamming weight.
The generated formula is
\[
F_{\mathrm{SCA}} =
F_{\mathrm{pt}} \land F_{\mathrm{state}} \land F_{\mathrm{key\_diff}} \land F_{\mathrm{leak}},
\]
encoding the plaintext, symbolic updates, key disequality, and leakage relation. Each execution evolves as
\[
S_t^{A/B}
=
\phi\left(
S_{t-1}^{A/B} \oplus \rho_t\left(K^{A/B}\right)
\right),
\]
with $A$ and $B$ denoting the two executions, $\rho_t$ the cycle-dependent key rotation, and $\phi$ the optional substitution layer. XOR, AND, equivalence, and disequality relations are Tseitin-encoded in DIMACS format.

For Hamming-weight leakage, a popcount circuit constrains the two leakage vectors to be equal or unequal. The reported CNFs use non-equality at the selected check cycle, so the solver searches for distinct key/state assignments that follow the same plaintext-driven update but differ in modeled leakage. Thus, leakage observations become hard CNF constraints solved by CDCL or QGCL, with candidates verified against cryptographic and leakage constraints.

\subsection{Grover Search}

Grover's algorithm provides a quadratic quantum speedup for unstructured search~\cite{grover1996fast}. Given an oracle $O$ that marks solutions in a finite domain $\mathcal{X}$ of size $N$,
\[
O\lvert x \rangle = (-1)^{f(x)} \lvert x \rangle,
\qquad
f(x) \in \{0,1\},
\]
Grover's algorithm repeatedly applies the unitary $G = DO$, where $D$ is the inversion-about-the-mean, or diffusion, operator. These iterations rotate amplitude toward the solution subspace. When the number of solutions $K = \lvert \{x : f(x) = 1\} \rvert$ is known, the optimal number of iterations scales as $\Theta(\sqrt{N/K})$, after which measuring the state yields a solution with high probability.

In SAT, one may define $\mathcal{X} = \{0,1\}^n$ and $f(x) = F(x)$, leading to a quantum search over all assignments. However, this direct application requires an oracle that coherently evaluates the entire CNF and a diffusion operator over all $n$ variables plus $m$ clauses; \textit{both the required number of iterations and the circuit size scale poorly with problem size.}

\subsection{Quantum Approaches to SAT Problems}

A substantial body of work investigates quantum algorithms for SAT and related NP-hard problems. Direct applications of Grover search treat each assignment $x \in \{0,1\}^n$ as a basis state and implement an oracle that evaluates the entire formula $F(x)$; repeating Grover iterations then yields a satisfying assignment (if one exists) with high probability in $O(\sqrt{2^n})$ oracle calls \cite{fernandes2019using, alasow2022quantum}. In addition, prior works encode SAT or MaxSAT as an energy landscape and use variational circuits or quantum annealing to minimize a cost Hamiltonian~\cite{kruger2020quantum, battaglia2005optimization}. For example, recent work on QAOA for random $k$-SAT analytically characterizes its success probability at the satisfiability threshold and reports regimes where QAOA outscales a specialized classical solver such as WalkSATlm~\cite{boulebnane2024solving}. These results are encouraging, but they assume global access to the entire instance and deep circuits whose width grows with the full number of variables, as these conditions are difficult to meet in realistic, large structured cryptanalytic CNFs.

Several works also use quantum subroutines to accelerate search procedures that remain classical at a higher level. Parallel and distributed Grover-based SAT solvers use entangled ancillas and teleportation to evaluate many clauses across quantum nodes, reducing oracle depth and spreading the qubit burden~\cite{lin2024parallel}. Montanaro's quantum walk algorithm, by contrast, accelerates abstract backtracking trees and yields a quadratic improvement in the number of explored nodes for DPLL-style search~\cite{montanaro2015quantum}. Measurement-driven $k$-SAT algorithms use weak or generalized measurements, often with Zeno-type dynamics, to steer the system toward satisfying subspaces. For small 3-SAT encodings, numerical studies report runtimes of the form $\lambda^n$ with $\lambda \lesssim \sqrt{2}$~\cite{zhang2025solving}. These results show that the general idea of accelerating SAT-style search with quantum subroutines has clear precedents. What they do not provide, however, is a concrete interface to the mechanisms that dominate \emph{modern} solver performance: conflict analysis, conflict-local subformula extraction, and VSIDS-style branching. In addition, all these methods treat the quantum algorithm as a stand-alone solver
that replaces the classical engine on the full instance, and typically require circuits whose size still scales with the global
problem dimension.

HyQSAT~\cite{tan2023hyqsat} is the closest solver-level hybrid predecessor. It combines a D-Wave quantum annealer with a CDCL solver by mapping selected clauses to an Ising model and using annealing outcomes to estimate clause satisfaction probabilities. These estimates guide clause prioritization and branching. While HyQSAT reports speedups on some 3-SAT benchmarks, it also exposes limitations of QA-based hybrid solvers: (i) long and topology-dependent embedding times, which dominate runtime for large formulas; (ii) sensitivity to hardware noise and coefficient tuning; and (iii) outputs that are heuristic scores rather than exact evaluations of a CNF subformula.

QGCL addresses a different gap. Instead of quantumizing the full instance, accelerating an abstract backtracking tree divorced from solver heuristics, or relying on
quantum annealing over a global Ising embedding, we integrate bounded Grover calls directly into a practical CDCL loop. Conflict analysis identifies when to call the quantum layer, Breadth-First Search (BFS) -based clause expansion defines the subproblem, and the returned samples are converted into concrete VSIDS activity and polarity updates. Because subproblems stay relatively small, qubit needs and circuit depth remain modest and do not scale with the full CNF, avoiding the high embedding cost. This engineering-level integration with conflict analysis and branching heuristics is the main point of departure from prior quantum SAT methods.

\begin{figure}[t!]
    \centering
    \includegraphics[width=\linewidth]{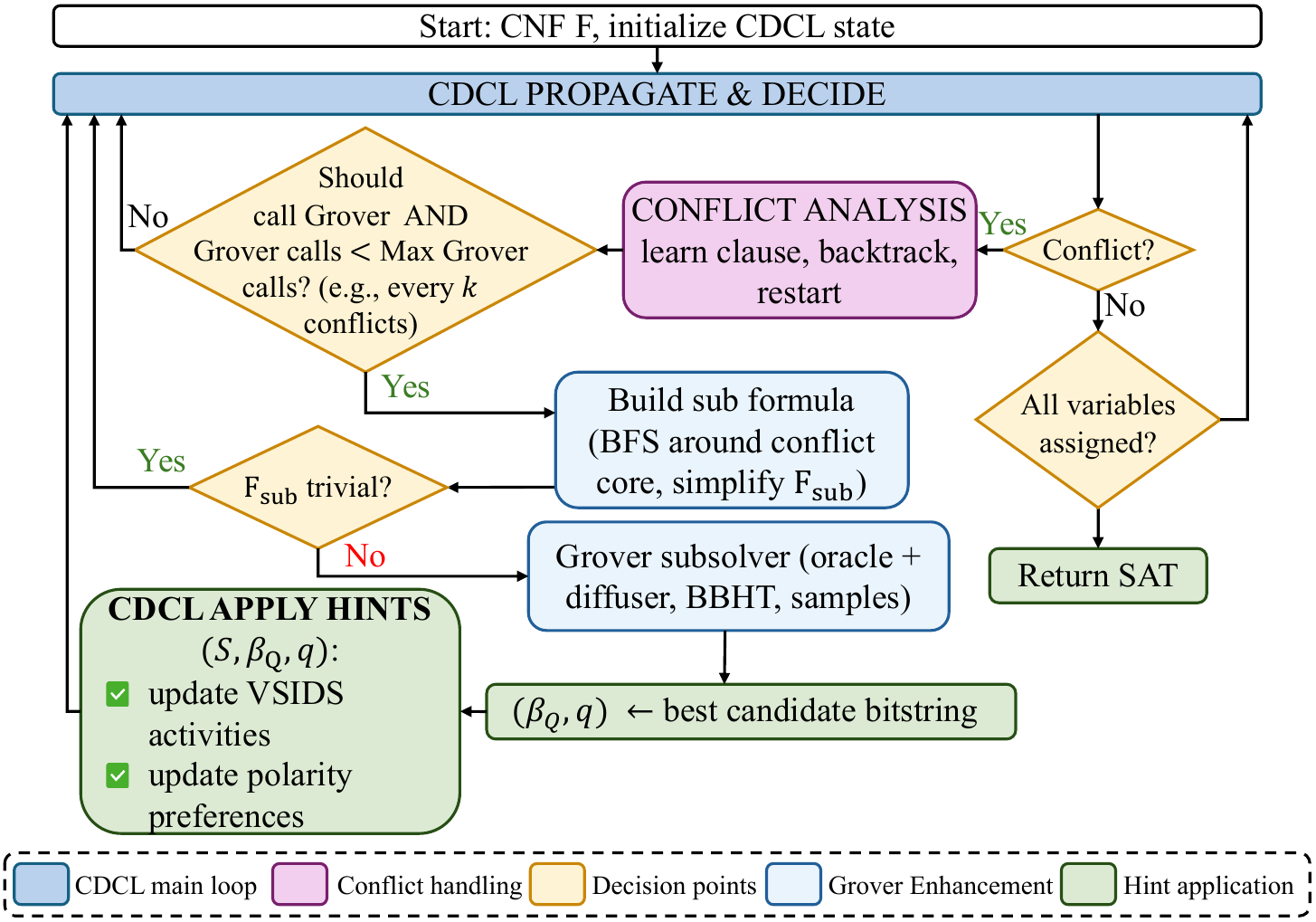}
    \caption{High-level control flow of the proposed QGCL solver. A standard CDCL loop performs propagate-decide steps and conflict analysis. After every $k$ conflicts (subject to a global call budget), the solver optionally extracts a local subformula $F_{\text{sub}}$ around the current conflict core, simplifies it under the partial assignment, and, if non-trivial, invokes a Grover subsolver (oracle + diffuser, BBHT schedule, sampling). The resulting heuristic pair $(\beta_Q, q)$ is fed back to CDCL to update VSIDS activities and polarity preferences; all SAT/UNSAT decisions remain purely classical.
}
    \label{fig:QGCLOverview}
\vspace{-15pt}
\end{figure}

\section{Hybrid Grover-CDCL Framework}

QGCL keeps a conventional conflict-driven clause learning solver as the decision procedure. The quantum layer is invoked only on subformulas extracted during search and influences only heuristic state. This section first gives a workflow view and a simplified runtime model, then details the Grover circuit and its embedding into CDCL.

\subsection{Workflow overview}
Before detailing the oracle, it is useful to summarize the control flow of QGCL shown in Figure~\ref{fig:QGCLOverview}. Each quantum interaction has four stages: (1) \emph{conflict detection}, where CDCL reaches a conflict and performs its standard analysis; (2) \emph{subformula extraction}, where a high-activity seed clause is expanded by BFS over the clause-variable incidence graph to collect clauses near the conflict core, then simplified under the current trail to obtain the bounded CNF $F_{\text{sub}}$ passed to Grover; (3) \emph{Grover search}, where a Boyer-Brassard-H{\o}yer-Tapp (BBHT)-style schedule~\cite{boyer1998tight} is run on $F_{\text{sub}}$ and returns sampled bitstrings together with a quality estimate; and (4) \emph{heuristic update}, where the best local assignment is translated into activity and polarity updates for the next CDCL decisions. The global clause database, restart logic, and proof-producing reasoning remain classical throughout.

For a selected subproblem with $n_{\text{sub}}$ variables and $m_{\text{sub}}$ clauses, we define the Grover budget as
\[
B = n_{\text{sub}} + m_{\text{sub}}.
\]
In the logical circuit model used here, the corresponding oracle width is
\[
W_{\text{logical}} = n_{\text{sub}} + m_{\text{sub}} + 1 = B + 1,
\]
where the extra qubit is the formula flag; backend-specific decompositions may introduce temporary work qubits. Thus $B$ is not identical to the eventual physical qubit count, but it is a clear proxy for logical circuit width and oracle size.

\subsection{QGCL Runtime Model}
Let $G$ be the number of Grover calls triggered during one CDCL run. For the $t$th call, let the simplified subformula contain $n_t$ variables, $m_t$ clauses, and $K_t$ satisfying assignments, with budget $B_t=n_t+m_t$. A first-order cost decomposition is
\[
C_t^{\text{call}} = C_t^{\text{ext}}(B_t) + C_t^{\text{Grover}}(n_t,m_t,K_t) + C_t^{\text{fb}},
\]
where $C_t^{\text{ext}}$ is the BFS extraction/simplification cost and $C_t^{\text{fb}}$ is the cost of updating heuristic state. Under BBHT~\cite{boyer1998tight}, the expected quantum part is
\[
C_t^{\text{Grover}} \approx \gamma \, c_{\text{iter}}(n_t,m_t)\sqrt{\frac{2^{n_t}}{\max(K_t,1)}},
\]
with $\gamma$ absorbing repeated sampling and classical checking of the returned bitstrings. In our construction, $c_{\text{iter}}$ includes clause-ancilla computation, formula-flag aggregation, uncomputation, and the diffuser, and is bounded by the chosen budget.

This quantum-cost term should be interpreted as a wall-clock cost when QGCL is executed on a simulator, fake backend, or real backend. Because $K_t$ is generally unknown for extracted sub-CNFs, BBHT may require multiple randomized guesses of the Grover iteration count before a useful candidate appears. Each guess incurs circuit construction or transpilation, quantum execution or simulation, shot collection, and classical verification of the returned bitstrings. When $K_t=0$, when there are many unevenly amplified satisfying assignments, or when noise hides the amplified states, BBHT can consume additional attempts without producing strong guidance. Thus the practical runtime depends not only on $G$ and $B_t$, but also on the number of BBHT attempts, the shot count, and the backend noise/simulation model.

To connect this quantum cost to CDCL behavior, suppose the current classical heuristic places probability mass $\mu_t$ on a locally productive subset $\mathcal{B}_t \subseteq \{0,1\}^{n_t}$ of assignments. After $r$ Grover iterations, amplitude amplification changes this mass to
\[
\widetilde{\mu}_t(r)=\sin^2\!\left((2r+1)\arcsin\sqrt{\mu_t}\right).
\]
QGCL does not branch on full assignments directly; instead, it extracts variable-wise marginals $\widehat{\pi}_{t,v}$ from the amplified samples and mixes them with the current CDCL branching preferences,
\[
\pi'_{t,v}=(1-\eta_t)\pi_{t,v}+\eta_t\widehat{\pi}_{t,v},
\]
where $\eta_t$ is reduced when the returned violation score is high. If $\Lambda_t$ denotes the local conflict density (the expected number of future conflicts reachable from the current subtree), then a simple proxy for the expected conflict reduction is
\[
\Delta_t \propto \Lambda_t\bigl(\widetilde{\mu}_t-\mu_t\bigr).
\]
This yields the coarse hybrid-runtime model
\[
\mathbb{E}[T_{\text{QGCL}}]
\approx
T_{\text{CDCL}}
+\sum_{t=1}^{G} C_t^{\text{call}}
-\bar c_{\text{conf}}\sum_{t=1}^{G}\Delta_t,
\]
where $\bar c_{\text{conf}}$ is the average cost per CDCL conflict. The model predicts a useful operating regime when each Grover call is cheap enough to satisfy the budget, yet informative enough that the induced increase in productive-branch mass produces a non-negligible drop in future conflicts. This is exactly the regime explored in our budget and call-count sweeps.

\subsection{Grover circuit construction}\label{Subsec:GroverCircuit}

For a subformula $F_{\text{sub}}$ over $n_{\text{sub}}$ variables and $m_{\text{sub}}$ clauses, the quantum back-end implements a Grover-style search over assignments $x \in \{0,1\}^{n_{\text{sub}}}$. The construction follows a standard compute-phase-uncompute pattern, specialized to CNF~\cite{fernandes2019using}.

The circuit uses three distinct qubit registers. The first is a variable register $\mathcal{H}_X$ consisting of $n_{\text{sub}}$ qubits, which encodes bit strings $x \in \{0,1\}^{n_{\text{sub}}}$. The second register is a set of clause ancilla qubits $c_1, \ldots, c_{m_{\text{sub}}}$, each of which is a single qubit associated with an individual clause. Finally, a single formula flag qubit $f$ is employed to aggregate the information from all clause ancillas.

We initialize all ancillas in $\ket{0}$. The variable register is prepared
in the uniform superposition
\[
\ket{s}
= H^{\otimes n_{\text{sub}}} \ket{0^{n_{\text{sub}}}}
= \frac{1}{\sqrt{N}} \sum_{x \in \{0,1\}^{n_{\text{sub}}}} \ket{x},
\qquad
N = 2^{n_{\text{sub}}}.
\]

For each clause $C_j = \ell_{j,1} \vee \cdots \vee \ell_{j,p_j},$ $\ell_{j,i} \in \{ x_k, \neg x_k \},$ we implement a reversible circuit that sets the ancilla qubit $c_j = 1$ exactly when the clause is violated. A convenient way to encode this is to interpret each literal $\ell_{j,i}$ as the condition that “this literal is false.” Concretely, for a positive literal $x_k$, the condition “false” corresponds to $x_k = 0$, while for a negated literal $\neg x_k$, it corresponds to $x_k = 1$. We then apply a multi-controlled $X$ gate on $c_j$, conditioned on all literals in the clause being false. In practice, this is implemented by placing $X$ gates around the control on $x_k$ whenever the active (triggering) value is $0$, and using standard controls without surrounding $X$ gates when the active value is $1$.
After this step, $c_j = 1$ if and only if $C_j$ is unsatisfied under the current assignment encoded in $\mathcal{H}_X$.

% \paragraph{Formula flag and phase oracle.}
To flag satisfying assignments, we aggregate the clause ancillas via an AND tree:
\[
f \leftarrow \bigwedge_{j=1}^{m_{\text{sub}}} \neg c_j .
\]

Equivalently, one can first compute an OR-of-violations
\[
v = \bigvee_j c_j
\]
and then set $f = \neg v$. The resulting flag qubit satisfies
\[
f = 1 \;\Longleftrightarrow\; F_{\text{sub}}(x) = 1 .
\]

We then apply a conditional phase flip on $\mathcal{H}_X$ controlled by $f$:
\[
\ket{x}\ket{f=1} \mapsto -\ket{x}\ket{f=1},
\qquad
\ket{x}\ket{f=0} \mapsto \ket{x}\ket{f=0}.
\]

\begin{figure}
    \centering
    \includegraphics[width=0.6\linewidth]{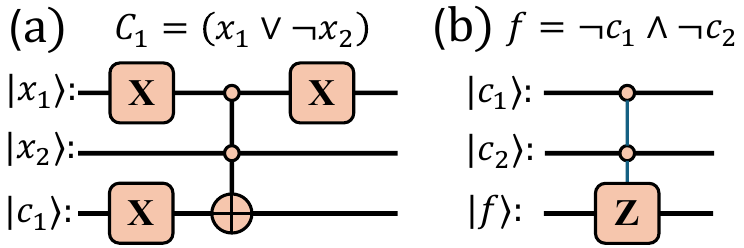}
    \caption{Circuit sketch for SAT-oracle building blocks.
(a) Clause-violation ancilla $c_1$ for the clause $C_1 = (x_1 \lor \neg x_2)$. $X$ gates on the positive literal $x_1$ retarget the control so that the multi-controlled $X$ on $c_1$ fires only when both literals in $C_1$ are false (clause violated), i.e., when $x_1 = 0$ and $x_2 = 1$. Thus $c_1 = 1$ encodes $C_1$ violated.
(b) Two such clause-violation bits $c_1, c_2$ are combined with a multi-controlled Toffoli to implement
$
f \leftarrow \neg c_1 \wedge \neg c_2,
$
the CNF flag that is $1$ iff no clause is violated. A controlled-$Z$ on $f$ applies the Grover phase to satisfying assignments. All ancillas are later uncomputed.
}
    \label{fig:GroverCircuit}
    \vspace{-15pt}
\end{figure}

This realizes a unitary
\[
U_{F_{\text{sub}}}\ket{x}\ket{0^A}
= (-1)^{F_{\text{sub}}(x)} \ket{x}\ket{0^A},
\]
where $\lvert 0^A \rangle$ denotes all ancillas (clause ancillas and the flag qubit) initialized to zero. Finally, we uncompute the AND tree and each clause circuit to restore all ancillas to $\ket{0}$, leaving only the phase kickback on the variable register. An illustrative example of the oracle building block is provided in Figure~\ref{fig:GroverCircuit}.

% \paragraph{Diffuser.}
The diffusion operator acts only on the variable register:
\[
D
= 2\ket{s}\!\bra{s} - I
= H^{\otimes n_{\text{sub}}}
\bigl(2\ket{0^{n_{\text{sub}}}}\!\bra{0^{n_{\text{sub}}}} - I\bigr)
H^{\otimes n_{\text{sub}}}.
\]

One Grover iteration is the composition
$
G = D U_{F_{\text{sub}}}.
$

When the number of satisfying assignments
$
K = \lvert \{ x : F_{\text{sub}}(x) = 1 \} \rvert
$
is unknown, we employ a BBHT-style schedule~\cite{boyer1998tight}: in each attempt, choose a random iteration count $r \sim \mathrm{Unif}\{0, \ldots, M-1\}$, apply $r$ iterations of $G$, measure, and verify the outcome classically. The parameter $M$ increases multiplicatively to $O(\sqrt{N})$, ensuring the expected query complexity $O(\sqrt{N/K})$ when $K \ge 1$.
In our implementation, each Grover attempt produces many bitstrings in a single shot-batched run; only the most frequent candidates are passed to the classical checker.

\subsection{QGCL Framework}

We now detail the CDCL-side mechanics that surround the Grover call. When the conflict counter reaches a call point, the solver first forms a queue of candidate clauses by selecting a seed clause from a high-activity region and performing a breadth-first expansion over the clause-variable incidence graph~\cite{moore1959shortest, lee1961algorithm}. Clauses already satisfied by the current trail are discarded, falsified literals are removed, and the remaining set is trimmed until the budget constraint is met. The surviving variables are then remapped densely to $\{1,\ldots,n_{\text{sub}}\}$ so that the extracted CNF can be passed directly to the Grover circuit generator.

\begin{figure*}[t]
    \centering
    \includegraphics[width=0.95\linewidth]{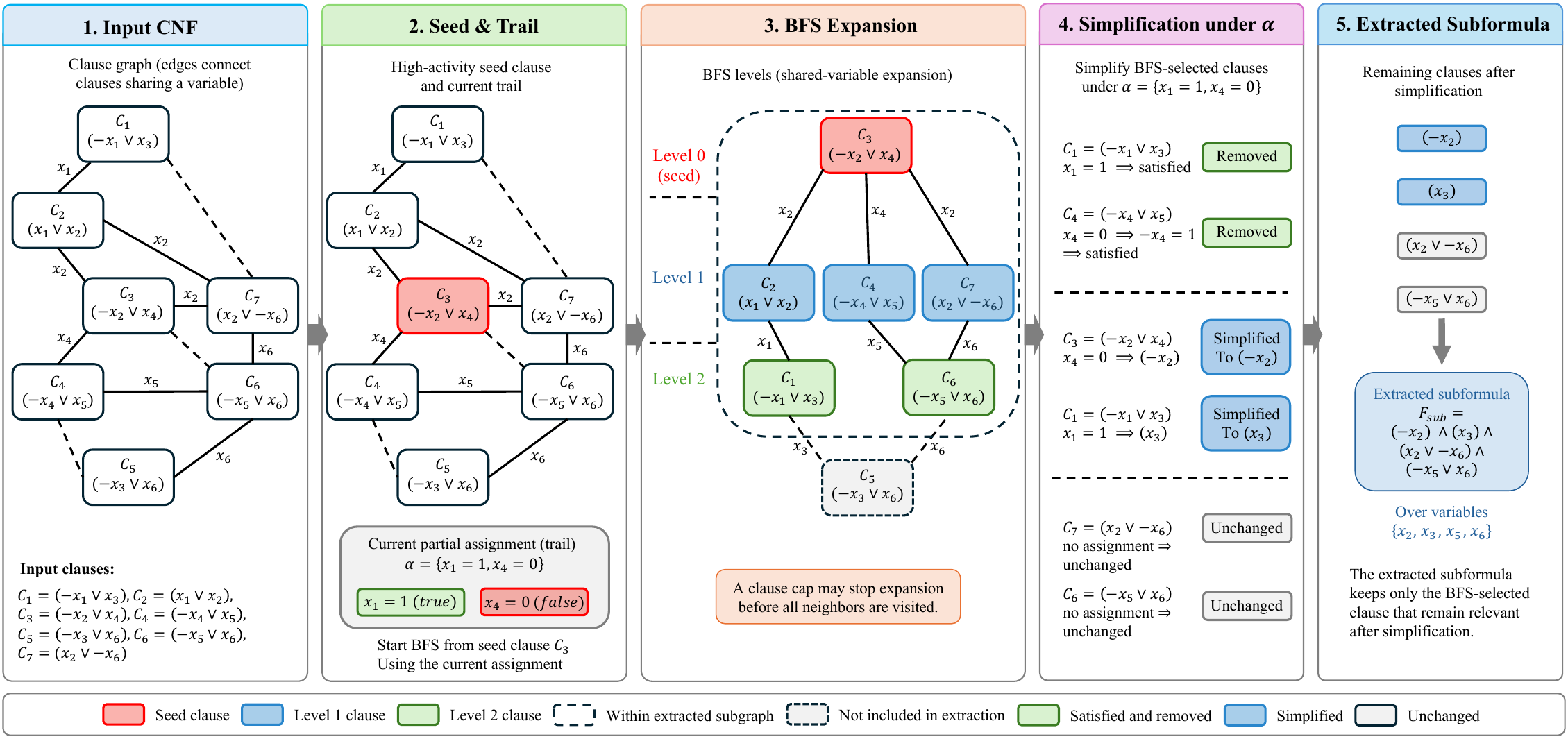}
    \caption{Schematic illustration of BFS-based subformula extraction and simplification from an input CNF formula. Each clause $C_i$ is represented as a node in a clause graph, and an edge between two clauses indicates that they share the variable written on the edge. Starting from the high-activity seed clause $C_3 = (-x_2 \vee x_4)$ and the current partial assignment $\alpha = \{x_1 = 1, x_4 = 0\}$, BFS expands through shared-variable neighbors to collect a local subgraph: $C_3$ at Level 0, $C_2, C_4, C_7$ at Level 1, and $C_1, C_6$ at Level 2. The dashed boundary marks the extracted subgraph, while $C_5$ is shown outside the boundary because the clause cap stops the expansion before all reachable neighbors are included. The selected clauses are then simplified under $\alpha$: $C_2$ and $C_4$ are satisfied and removed, $C_3$ reduces to $(-x_2)$, $C_1$ reduces to $(x_3)$, and $C_7, C_6$ remain unchanged. The resulting extracted subformula is $F_{\mathrm{sub}} = (-x_2) \wedge (x_3) \wedge (x_2 \vee -x_6) \wedge (-x_5 \vee x_6)$, over the remaining variables $\{x_2, x_3, x_5, x_6\}$.}
    \label{fig:BFS}
    \vspace{-14pt}
\end{figure*}

Figure~\ref{fig:BFS} shows a schematic example of this extraction process. The example highlights why BFS expansion is useful: starting from one conflict-relevant clause, the queue quickly collects neighboring clauses that share variables with the seed, thereby preserving a localized slice of the conflict structure. Simplification under the trail then removes already-resolved context, leaving a smaller CNF that is both more informative than a single clause and still small enough for the quantum budget. If simplification collapses the candidate set to a trivial formula, or if the selected clauses cannot be fit inside the budget, the solver skips the quantum call and continues with ordinary CDCL.

\begin{figure}
    \centering
    \includegraphics[width=1\linewidth]{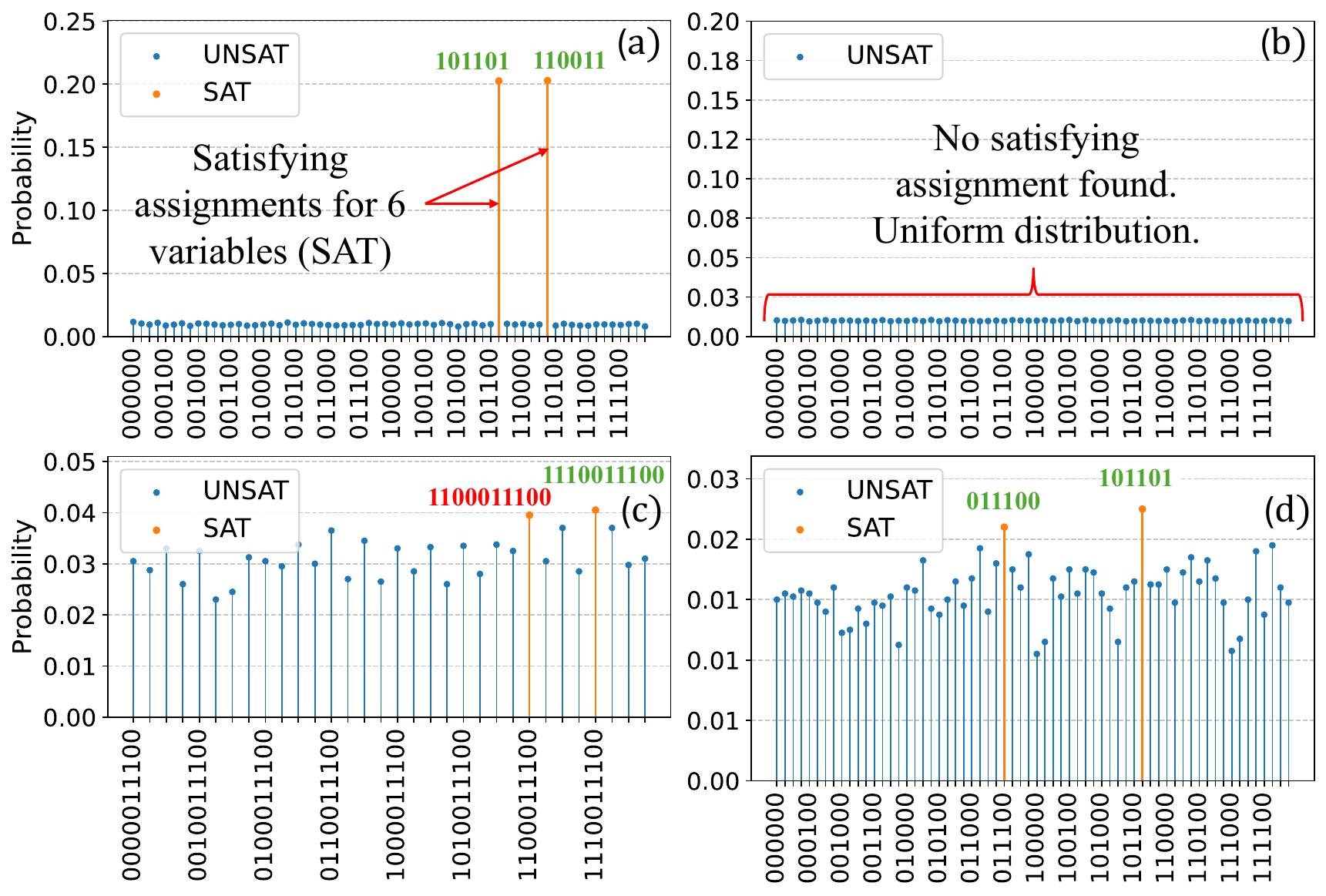}
    \caption{
Grover-subsolver output distributions for small CNF formulas used to validate the local oracle and sampling loop. For an extracted CNF with $n_{\mathrm{sub}}$ variables and $m_{\mathrm{sub}}$ clauses, the oracle uses $n_{\mathrm{sub}}+m_{\mathrm{sub}}+1$ qubits. Panels (a,b) show ideal simulations: satisfiable cases amplify marked assignments, while unsatisfiable cases remain near-uniform. Panels (c,d) show noisy Fake Prague simulations, where the correct assignment is often still among the highest-probability candidates despite noise. 
    }
    \label{fig:GroverSAT}
    \vspace{-15pt}
\end{figure}

Given a non-trivial subformula $F_{\text{sub}}$, the classical driver invokes the Grover subsolver. The subsolver constructs the phase oracle $U_{F_{\text{sub}}}$ and diffuser $D$ as described in Section~\ref{Subsec:GroverCircuit}, executes several BBHT attempts with randomly chosen iteration counts $r$, and measures the variable register. From these samples, QGCL retains only a small set of high-frequency candidate bitstrings and classically scores them by the fraction of subclauses they violate. The best candidate defines a partial assignment $\beta_Q$ on the variables of $F_{\text{sub}}$ together with a violation score $q \in [0,1]$, where $q$ is the violated-clause fraction; lower is better. Figure~\ref{fig:GroverSAT} provides diagnostic small-CNF examples of this Grover subsolver behavior under both ideal and noisy simulation. These examples are not AES-scale key-recovery instances; they verify how the local Grover oracle amplifies satisfying candidates, how an UNSAT subformula produces an uninformative near-uniform histogram, and how noisy fake-backend runs can still yield a useful top-ranked candidate for CDCL guidance.

The pair $(\beta_Q,q)$ is fed back only through heuristic state. When $q$ is small, QGCL treats $\beta_Q$ as low-violation, near-satisfying local guidance: variables that occur frequently in $F_{\text{sub}}$ receive polarity preferences consistent with $\beta_Q$ and their VSIDS activities are increased. When $q$ is large, the subformula is interpreted as a conflict-rich region; activities of its variables are boosted, but little or no polarity bias is applied. This violation-score-aware feedback also provides robustness to noisy quantum samples, because flat or inconsistent histograms naturally lead to weak updates. The Fake Prague examples in Figure~\ref{fig:GroverSAT}(c)-(d) clarify the practical meaning of this robustness. QGCL does not require every amplified state to be correct: it keeps only the most frequent candidates, evaluates them against $F_{\text{sub}}$, and uses the resulting score only to bias future CDCL decisions. A wrong secondary peak can weaken the heuristic signal or increase the number of BBHT attempts, but it cannot certify an invalid assignment or change the logical SAT/UNSAT result.

The interaction between the classical and quantum components can be summarized in Algorithm~\ref{alg:hybrid-grover-cdcl}.

\begin{algorithm}[t]
\caption{Proposed QGCL Algorithm}
\label{alg:hybrid-grover-cdcl}
% \small
% \scriptsize
\footnotesize
\begin{algorithmic}[1]
\Require CNF formula $F$ over variables $X = \{x_1, \ldots, x_n\}$
\Ensure \textsc{SAT} with model, or \textsc{UNSAT}

\State Initialize CDCL state $S$ for $F$ \Comment{watched literals, activities, etc.}
\State $\textit{grover\_calls} \gets 0$

\While{\textbf{true}}
    \State $\textit{confl} \gets \Call{CDCL\_Propagate}{S}$
    \If{$\textit{confl} \neq \bot$}
        \State $(\textit{cl\_learnt}, \textit{lvl}) \gets \Call{CDCL\_Analyze}{S, \textit{confl}}$
        \State \Call{CDCL\_LearnAndBacktrack}{$S, \textit{cl\_learnt}, \textit{lvl}$}
        \If{\Call{ShouldRestart}{$S$}}
            \State \Call{CDCL\_Restart}{$S$}
        \EndIf
        \If{\Call{ShouldCallGrover}{$S$} \textbf{and} $\textit{grover\_calls} < \textit{MaxGroverCalls}$}
            \State $F_{\text{sub}} \gets \Call{BuildSubformula}{F, S}$ 
            \Comment{BFS over clause graph, simplification}
            \If{$F_{\text{sub}} \neq \bot$}
                \State $\textit{grover\_calls} \gets \textit{grover\_calls} + 1$
                \State $(\beta_Q, q) \gets \Call{GroverSubSolve}{F_{\text{sub}}}$
                \Comment{quantum back-end (oracle + diffuser)}
                \State \Call{CDCL\_ApplyHints}{$S, \beta_Q, q$}
                \Comment{update polarities / activities only}
            \EndIf
        \EndIf
    \Else
        \If{\Call{AllVariablesAssigned}{$S$}}
            \State \Return $(\textsc{SAT}, \Call{ExtractModel}{S})$
        \EndIf
        \State $\ell \gets \Call{CDCL\_PickBranchLiteral}{S}$
        \If{$\ell = \bot$}
            \State \Return $(\textsc{SAT}, \Call{ExtractModelWithDefaults}{S})$
        \EndIf
        \State \Call{CDCL\_NewDecisionLevel}{$S$}
        \State \Call{CDCL\_Enqueue}{$S, \ell$}
    \EndIf
\EndWhile
\end{algorithmic}
\end{algorithm}

In Algorithm~\ref{alg:hybrid-grover-cdcl}, the procedure
\textsc{ShouldCallGrover} implements a simple conflict-based sampling policy, for example, invoking the quantum subsolver every $k$ conflicts up to a fixed overall budget. The routine \textsc{BuildSubformula} encapsulates the clause-queue generation and simplification process described above. The procedure \textsc{GroverSubSolve} executes the Grover circuit construction of Section \ref{Subsec:GroverCircuit} and returns an assignment together with an associated violation score measuring the fraction of subclauses it violates. Finally, \textsc{CDCL\_ApplyHints} modifies only heuristic state within the CDCL solver, such as VSIDS activity scores and polarity preferences, without adding or removing clauses from the clause database.

For each subformula $F_{\text{sub}}$, the Grover routine outputs a distribution over bitstrings together with $(\beta_Q, q)$. Sharply peaked histograms are interpreted as low-violation candidate assignments, while nearly uniform histograms indicate weak or inconclusive guidance; in such cases, the corresponding hints are down-weighted through a large $q$.

QGCL modifies only heuristic state inside CDCL, specifically variable activities and polarity preferences. It never learns clauses from quantum output, never accepts a quantum assignment as a certificate, and never prunes branches without classical justification. Therefore, a returned \textsc{SAT} model is validated by classical CNF checking, and any \textsc{UNSAT} result follows solely from standard CDCL reasoning. The quantum layer can change the \emph{search order}, but not the logical meaning of the answer.

\vspace{-6pt}
\section{Experimental Setup}

We evaluate QGCL on controlled AES-oriented power side-channel CNFs generated by the workflow in Section~\ref{subsec:aes_sca_encoding}. The security target is AES-128 key recovery, while the reported benchmark files use a 100-bit symbolic key/state width as a scalable experimental proxy. The depth variable $T$ denotes the number of symbolic update/leakage cycles generated before the leakage relation is checked. Across $T\in\{2,4,8,12\}$, the CNF size grows from $9{,}249$ variables and $31{,}572$ clauses to $39{,}389$ variables and $137{,}712$ clauses. Unless a parameter is being swept explicitly, we use a default Grover budget of $B=20$ and a maximum Grover calls $C_{max}=15$ per SAT instance.

The baseline comparison uses the four instance sizes above with fixed $(B,\;C_{max})=(25,10)$. The call-count sweep fixes the smallest instance and varies the maximum number of Grover calls in $\{5,10,15,20\}$. The budget sweep on the same instance fixes 10 calls and varies $B\in \{5,10,15,20,25\}$. Finally, we assess the impact of noise on a noisy backend. Because $W_{\text{logical}}=B+1$ for our logical oracle, this range corresponds to roughly 6-26 logical qubits, excluding backend-specific work qubits introduced by gate decomposition.

\begin{table}[t]
\centering
\caption{Benchmark instances and Grover parameter settings used in the experiments.}
\label{tab:exp-summary}
\scriptsize
\setlength{\tabcolsep}{2.5pt}
\renewcommand{\arraystretch}{1.2}

\newcolumntype{C}[1]{>{\centering\arraybackslash}p{#1}}

\begin{tabular}{|C{0.20\columnwidth}|C{0.06\columnwidth}|C{0.25\columnwidth}|C{0.18\columnwidth}|C{0.14\columnwidth}|}
\hline
\textbf{Experiment} &
\textbf{$T$} &
\textbf{\makecell{Instance size\\$(n,m)$}} &
\textbf{\makecell{Grover\\budget $B$}} &
\textbf{\makecell{Max.\\calls}} \\
\hline

\multicolumn{5}{|l|}{\textit{Baseline comparison}} \\
\hline
\multirow{4}{0.20\columnwidth}{\centering Baseline\\vs.\\QGCL}
& 2  & $(9249,31572)$   & 20 & 15 \\
\cline{2-5}
& 4  & $(15277,52800)$  & 20 & 15 \\
\cline{2-5}
& 8  & $(27333,95256)$  & 20 & 15 \\
\cline{2-5}
& 12 & $(39389,137712)$ & 20 & 15 \\
\hline

\multicolumn{5}{|l|}{\textit{Parameter sweeps on the fixed $T=2$ instance}} \\
\hline
Call-count sweep &
2 &
$(9249,31572)$ &
20 &
\makecell[c]{$\{5,10,$\\$15,20\}$} \\
\hline
Budget sweep &
2 &
$(9249,31572)$ &
\makecell[c]{$\{5,10,15,$\\$20,25\}$} &
15 \\
\hline

\multicolumn{5}{|l|}{\textit{Noisy backend assessment}} \\
\hline
Baseline vs. QGCL &
2 &
$(9249,31572)$ &
10 &
8 \\
\hline

\end{tabular}
\vspace{-10pt}
\end{table}

The ideal-simulator experiments are carried out within the PennyLane software framework~\cite{bergholm2018pennylane}. Simulations are executed on an NVIDIA GeForce RTX~4090 GPU equipped with $24,564\,\mathrm{MiB}$ of memory, using the \texttt{lightning.gpu} statevector simulator provided by PennyLane. Measurement statistics for each Grover circuit are obtained from $2{,}000$ shots. To reduce the impact of statistical variability due to finite sampling and the randomized BBHT iteration schedule, each benchmark setting is independently evaluated 10 times using distinct solver seeds, and the reported results correspond to averages over these runs. 
We also report a noisy backend (Fake Prague 33 qubits) study for the $T=2$, $B=10$, $C_{max}=8$ setting. Those rows use a noisy fake backend rather than the ideal statevector simulator, so their wall time includes both QGCL control overhead and the cost of noisy quantum-circuit simulation. Table~\ref{tab:exp-summary} summarizes the experimental setup employed in this study.

\vspace{-6pt}
\section{Results and Discussion}
\vspace{-6pt}
\subsection{Performance Analysis of the CDCL Baseline and QGCL}
\begin{figure}
    \centering
    \includegraphics[width=1\linewidth]{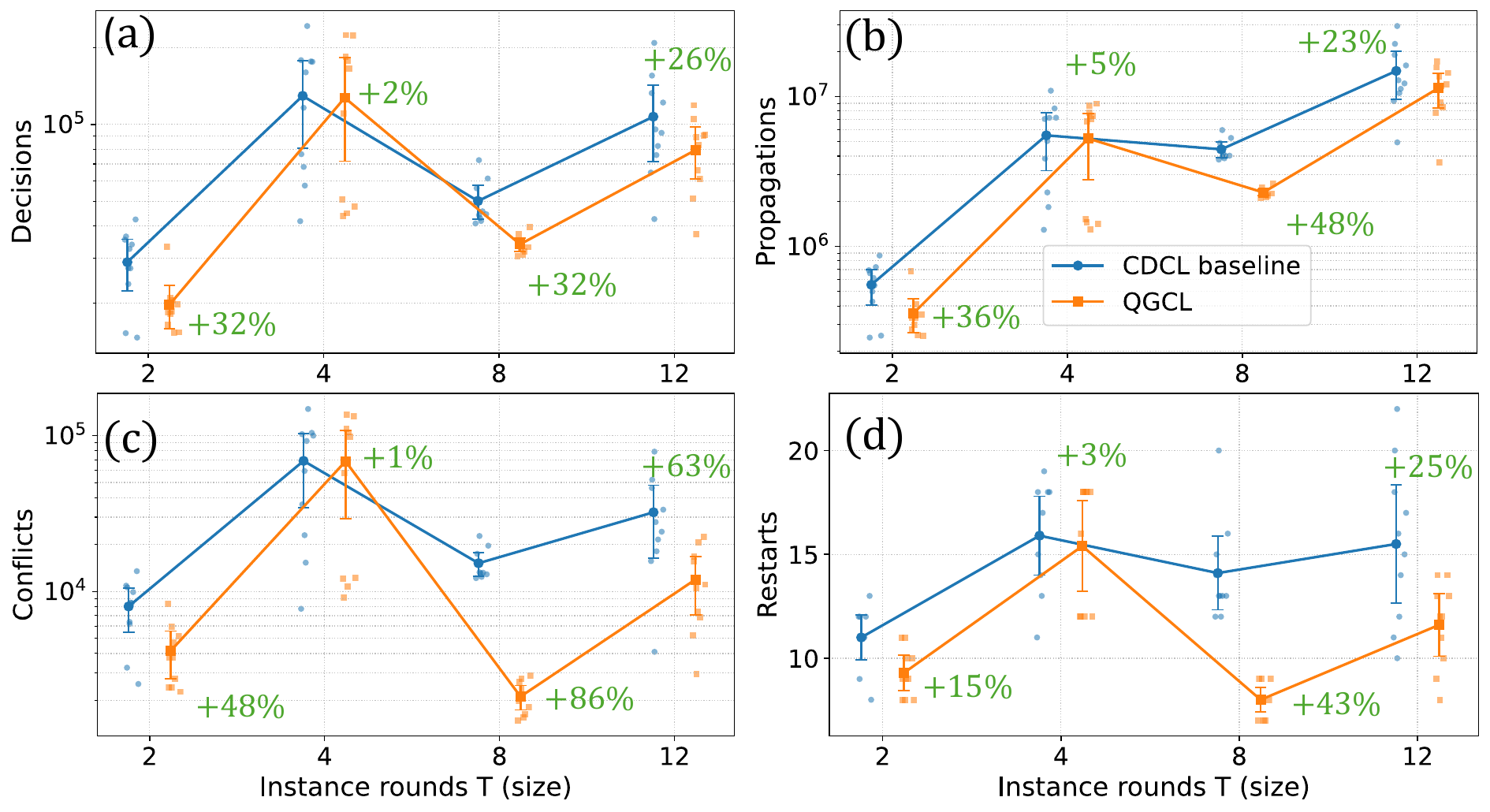}
    \caption{Comparison of the CDCL baseline (blue) and the QGCL hybrid solver (orange) on the benchmark family with $T \in \{2,4,8,12\}$. Panels show (a) decisions, (b) propagations, (c) conflicts, and (d) restarts. Points denote individual runs; solid markers and error bars show the mean and 95\% confidence interval across 10 seeds. Percentage labels indicate the relative improvement of QGCL over CDCL.
}
    \label{fig:HybridvsCDCL}
    \vspace{-15pt}
\end{figure}

In the first experiment, we compare the classical CDCL solver with the QGCL solver using four standard SAT statistics: restarts, conflicts, decisions, and propagations. We vary the instance parameter ($T\in{2,4,8,12}$), which controls the logical depth of the cryptographic encoding and the CNF size, Table~\ref{tab:exp-summary}. 
Figure~\ref{fig:HybridvsCDCL} reports CDCL statistics with per-run points and 95\% confidence intervals for the mean. Table~\ref{tab:qgcl_all_t} reports the corresponding means and standard deviations over 10 seeds.

The table and figure show that QGCL reduces CDCL search effort on three of the four depths. At $T=2$, QGCL lowers conflicts from $8{,}016.20$ to $4{,}155.50$ (about a $48\%$ reduction), with similar reductions in decisions and propagations. At $T=8$, the reduction is strongest: conflicts fall from $15{,}181.80$ to $2{,}115.70$ (about $86\%$), and restarts, decisions, and propagations also decrease. At $T=12$, QGCL again improves the search statistics, reducing conflicts by about $63\%$. The $T=4$ case is slightly favorable to QGCL the updated reductions are about $2\%$ in decisions, $5\%$ in propagations, $1\%$ in conflicts, and $3\%$ in restarts. 
This indicates that the Grover-guided hints are most useful when the extracted conflict cores contain enough structure to compensate for the overhead and randomness of the quantum subsolver. 

The wall-time column in Table~\ref{tab:qgcl_all_t} is also favorable to QGCL in the ideal-simulator experiments: the mean wall time is lower at all four depths, including $68.47$~s versus $70.32$~s at $T=4$ and $947.76$~s versus $4{,}348.46$~s at $T=12$. We interpret this optimistically, while still noting the large run-to-run variability at larger $T$ and the fact that these timings are from a simulator-based prototype. The main conclusion is that Grover-guided subformula search can reduce both CDCL search effort and simulated wall time when the extracted conflict cores are informative enough to compensate for oracle and BBHT overhead.

Overall, this shows that QGCL is most effective once instances increase in size and develop nontrivial conflict cores (e.g., $T\ge 8$). In this regime, Grover's guidance reduces search-tree exploration without changing the underlying decision procedure.

\subsection{Impact of the Maximum Number of Grover Calls on QGCL}

\begin{figure}
    \centering
    \includegraphics[width=1\linewidth]{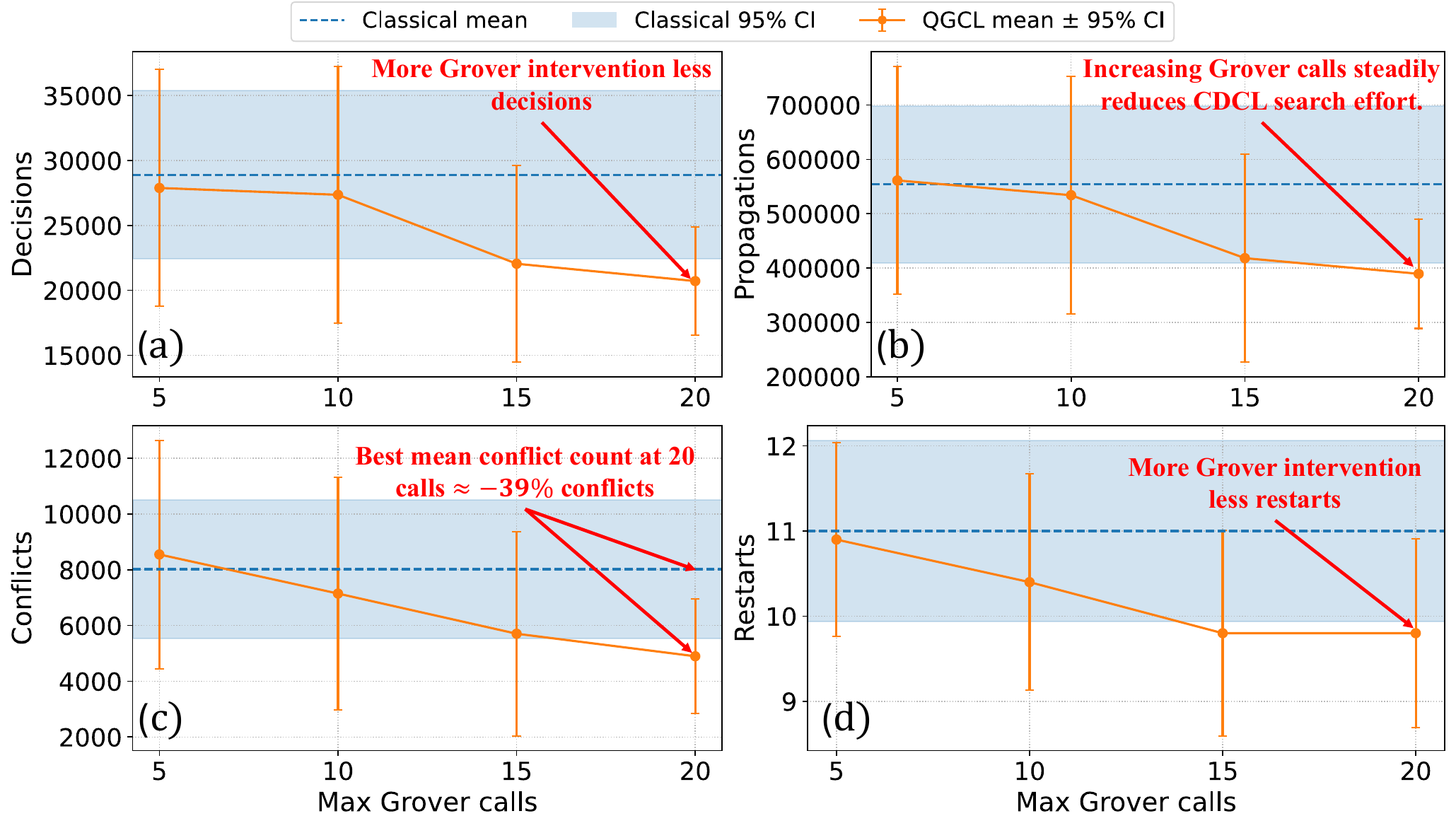}
    \caption{Performance of QGCL as a function of the allowed maximum number of Grover calls on a fixed benchmark instance. Panels report (a) decisions, (b) propagations, (c) conflicts, and (d) restarts. The dashed line and shaded band show the mean and 95\% confidence interval of the classical CDCL baseline; orange markers and error bars show the QGCL mean and 95\% confidence interval over 10 runs.}
    \label{fig:GroverCalls}
    \vspace{-15pt}
\end{figure}

In the second experiment, we fix the $T=2$ instance ($9{,}249$ variables and $31{,}572$ clauses) and sweep the maximum number of Grover calls $C_{max}$ allowed during CDCL. The Grover budget is fixed to $B=25$, so the sweep isolates how often the CDCL core can query the Grover subsolver. We run the QGCL solver 10 times and report the mean conflict count with a $95\%$ confidence band; the classical baseline is shown as a horizontal reference with its mean and CI. Figure~\ref{fig:GroverCalls} shows the per-metric trends, and Table~\ref{tab:t2_sweeps} gives the numerical values for the same $T=2$ sweep block.

The QGCL curve improves monotonically as max calls increases. With $C_{max}=5$, conflict counts are close to the baseline (Figure~\ref{fig:GroverCalls}c)), indicating that too few quantum interventions provide little benefit. At $C_{max}=10$, the conflict count drops and becomes comparable to the baseline, with overlapping CIs. 
At 15 and 20 calls, the hybrid solver’s mean conflict counts are lower than those of the classical solver, although the confidence intervals may still overlap.
The gain from 15 to 20 is smaller than from 5 to 10, suggesting diminishing returns. Similar trends hold for the decisions, propagations, and restarts statistics, see Figure~\ref{fig:GroverCalls} a), b), and d).

On the other hand, the wall-time value in Table~\ref{tab:t2_sweeps} are also favorable in the ideal simulator: $C_{\max}=20$ gives the lowest mean wall time in the call sweep ($3.92$~s versus $5.89$~s for CDCL). This indicates that, on the $T=2$ instance, the reduction in CDCL work can outweigh the cost of additional Grover calls when the call budget is sufficiently large.

Overall, these results show a positive trade-off: a non-negligible number of Grover calls is needed before quantum guidance has enough opportunities to influence CDCL, but once that threshold is reached, repeated local Grover searches systematically reduce search-tree exploration and improve the simulator wall time for this instance.

\begin{table*}[t]
\centering
\scriptsize
\setlength{\tabcolsep}{3pt}
\caption{QGCL versus classical CDCL across problem depths. Entries are mean $\pm$ sample standard deviation over available runs: 10 runs for $T=2,4,8$ and 9 runs for $T=12$. The lowest value in each comparable metric for each $T$ is bolded.}
\label{tab:qgcl_all_t}
\resizebox{\textwidth}{!}{%
\begin{tabular}{|c|c|c|c|c|c|c|c|c|}
\hline
\multicolumn{1}{|c|}{$T$} &
\multicolumn{1}{c|}{Solver} &
\multicolumn{1}{c|}{Vars} &
\multicolumn{1}{c|}{Clauses} &
\multicolumn{1}{c|}{Restarts} &
\multicolumn{1}{c|}{Conflicts} &
\multicolumn{1}{c|}{Decisions} &
\multicolumn{1}{c|}{Propagations} &
\multicolumn{1}{c|}{Wall time (s)} \\
\hline
2 & CDCL & 9,249 & 31,572
& $11.00 \pm 1.49$
& $8,016.20 \pm 3,460.64$
& $28,905.30 \pm 9,079.29$
& $553,733.80 \pm 201,932.30$
& $5.87 \pm 1.96$ \\
\hline
2 & QGCL & 9,249 & 31,572
& \textbf{9.30 $\pm$ 1.16}
& \textbf{4,155.50 $\pm$ 1,933.86}
& \textbf{19,679.50 $\pm$ 5,176.97}
& \textbf{356,261.50 $\pm$ 124,395.94}
& \textbf{3.90 $\pm$ 1.41} \\
\hline
4 & CDCL & 15,277 & 52,800
& $15.90 \pm 2.60$
& $68,864.80 \pm 47,213.22$
& $129,361.30 \pm 66,730.57$
& $5,504,358.60 \pm 3,166,241.05$
& $70.32 \pm 44.65$ \\
\hline
4 & QGCL & 15,277 & 52,800
& \textbf{15.40 $\pm$ 2.99}
& \textbf{68,455.50 $\pm$ 53,821.53}
& \textbf{127,190.60 $\pm$ 75,966.70}
& \textbf{5,242,834.70 $\pm$ 3,355,616.97}
& \textbf{68.47 $\pm$ 48.44} \\
\hline
8 & CDCL & 27,333 & 95,256
& $14.10 \pm 2.42$
& $15,181.80 \pm 3,587.89$
& $50,147.80 \pm 10,299.82$
& $4,439,207.80 \pm 721,100.07$
& $2,949.83 \pm 6,719.76$ \\
\hline
8 & QGCL & 27,333 & 95,256
& \textbf{8.00 $\pm$ 0.82}
& \textbf{2,115.70 $\pm$ 517.99}
& \textbf{33,951.90 $\pm$ 2,820.02}
& \textbf{2,288,233.90 $\pm$ 158,428.90}
& \textbf{2,768.25 $\pm$ 8,579.22} \\
\hline
12 & CDCL & 39,389 & 137,712
& $15.50 \pm 3.89$
& $32,219.30 \pm 21,697.63$
& $107,221.90 \pm 48,696.84$
& $14,816,735.50 \pm 7,173,855.31$
& $4,348.46 \pm 11,196.74$ \\
\hline
12 & QGCL & 39,389 & 137,712
& \textbf{11.60 $\pm$ 2.07}
& \textbf{11,936.00 $\pm$ 6,635.46}
& \textbf{79,315.00 $\pm$ 25,065.29}
& \textbf{11,357,000.00 $\pm$ 4,102,273.22}
& \textbf{947.76 $\pm$ 2,072.93} \\
\hline
\end{tabular}%
}
\end{table*}

\begin{table*}[t]
\centering
\scriptsize
\setlength{\tabcolsep}{3pt}
\caption{$T=2$ sweep experiments. Entries are mean $\pm$ sample standard deviation over 10 runs. The FakeBackend rows use the fake backend rather than the simulator. The lowest value in each sweep block is bolded.}
\label{tab:t2_sweeps}
\resizebox{\textwidth}{!}{%
\begin{tabular}{|c|c|c|c|c|c|c|c|}
\hline
\multicolumn{1}{|c|}{Experiment} &
\multicolumn{1}{c|}{Backend} &
\multicolumn{1}{c|}{Setting} &
\multicolumn{1}{c|}{Restarts} &
\multicolumn{1}{c|}{Conflicts} &
\multicolumn{1}{c|}{Decisions} &
\multicolumn{1}{c|}{Propagations} &
\multicolumn{1}{c|}{Wall time (s)} \\
\hline

\multirow{6}{*}{Budget sweep ($B$)}
& \multirow{6}{*}{Simulator}
& CDCL baseline
& $11.00 \pm 1.49$
& $8,016.20 \pm 3,460.64$
& $28,905.30 \pm 9,079.29$
& $553,733.80 \pm 201,932.30$
& $5.88 \pm 1.97$ \\
\cline{3-8}

& & QGCL, $B=5$
& $9.40 \pm 1.35$
& $4,197.80 \pm 2,814.97$
& $20,234.90 \pm 5,956.09$
& $369,942.60 \pm 147,586.83$
& $4.16 \pm 1.39$ \\
\cline{3-8}

& & QGCL, $B=10$
& $9.30 \pm 1.77$
& $4,581.00 \pm 4,087.91$
& $20,740.40 \pm 8,590.58$
& $384,826.30 \pm 197,086.62$
& $4.32 \pm 1.95$ \\
\cline{3-8}

& & QGCL, $B=15$
& $11.10 \pm 1.79$
& $9,914.50 \pm 6,368.09$
& $32,692.20 \pm 15,398.29$
& $665,870.20 \pm 348,480.18$
& $6.92 \pm 3.41$ \\
\cline{3-8}

& & QGCL, $B=20$
& \textbf{9.20 $\pm$ 1.32}
& \textbf{4,021.90 $\pm$ 2,985.33}
& \textbf{19,184.10 $\pm$ 6,338.39}
& \textbf{351,239.80 $\pm$ 164,068.07}
& \textbf{3.96 $\pm$ 1.48} \\
\cline{3-8}

& & QGCL, $B=25$
& $9.30 \pm 1.42$
& $4,471.50 \pm 3,166.93$
& $20,201.70 \pm 5,428.29$
& $372,205.00 \pm 151,683.65$
& $4.19 \pm 1.30$ \\
\hline

\multirow{5}{*}{Call sweep ($C_{\max}$)}
& \multirow{5}{*}{Simulator}
& CDCL baseline
& $11.00 \pm 1.49$
& $8,016.20 \pm 3,460.64$
& $28,905.30 \pm 9,079.29$
& $553,733.80 \pm 201,932.30$
& $5.89 \pm 1.89$ \\
\cline{3-8}

& & QGCL, $C_{\max}=5$
& $10.90 \pm 1.60$
& $8,546.30 \pm 5,725.27$
& $27,885.70 \pm 12,759.89$
& $561,225.40 \pm 292,871.50$
& $4.66 \pm 2.47$ \\
\cline{3-8}

& & QGCL, $C_{\max}=10$
& $10.40 \pm 1.78$
& $7,144.30 \pm 5,827.47$
& $27,358.20 \pm 13,805.44$
& $534,027.20 \pm 305,310.14$
& $4.48 \pm 1.03$ \\
\cline{3-8}

& & QGCL, $C_{\max}=15$
& \textbf{9.80 $\pm$ 1.69}
& $5,703.10 \pm 5,122.03$
& $22,047.40 \pm 10,600.52$
& $418,116.10 \pm 267,684.75$
& $4.32 \pm 1.81$ \\
\cline{3-8}

& & QGCL, $C_{\max}=20$
& \textbf{9.80 $\pm$ 1.55}
& \textbf{4,894.70 $\pm$ 2,871.43}
& \textbf{20,711.80 $\pm$ 5,792.36}
& \textbf{389,330.10 $\pm$ 140,560.57}
& \textbf{3.92 $\pm$ 1.45} \\
\hline

\multirow{2}{*}{Budget sweep ($B$)}
& \multirow{2}{*}{FakeBackend}
& CDCL baseline
& $11.00 \pm 1.49$
& $8,016.20 \pm 3,460.64$
& $28,905.30 \pm 9,079.29$
& $553,733.80 \pm 201,932.30$
& $6.11 \pm 2.02$ \\
\cline{3-8}

& & QGCL, $B=10$
& \textbf{9.70 $\pm$ 1.42}
& \textbf{4,459.80 $\pm$ 2,659.36}
& \textbf{21,062.60 $\pm$ 6,000.29}
& \textbf{372,370.20 $\pm$ 134,997.29}
& $80.81 \pm 10.06$ \\
\hline
\end{tabular}%
}
%\vspace{-10pt}
\end{table*}

\subsection{Impact of the Grover Qubit Budget on QGCL Performance}

\begin{figure}
    \centering
    \includegraphics[width=1\linewidth]{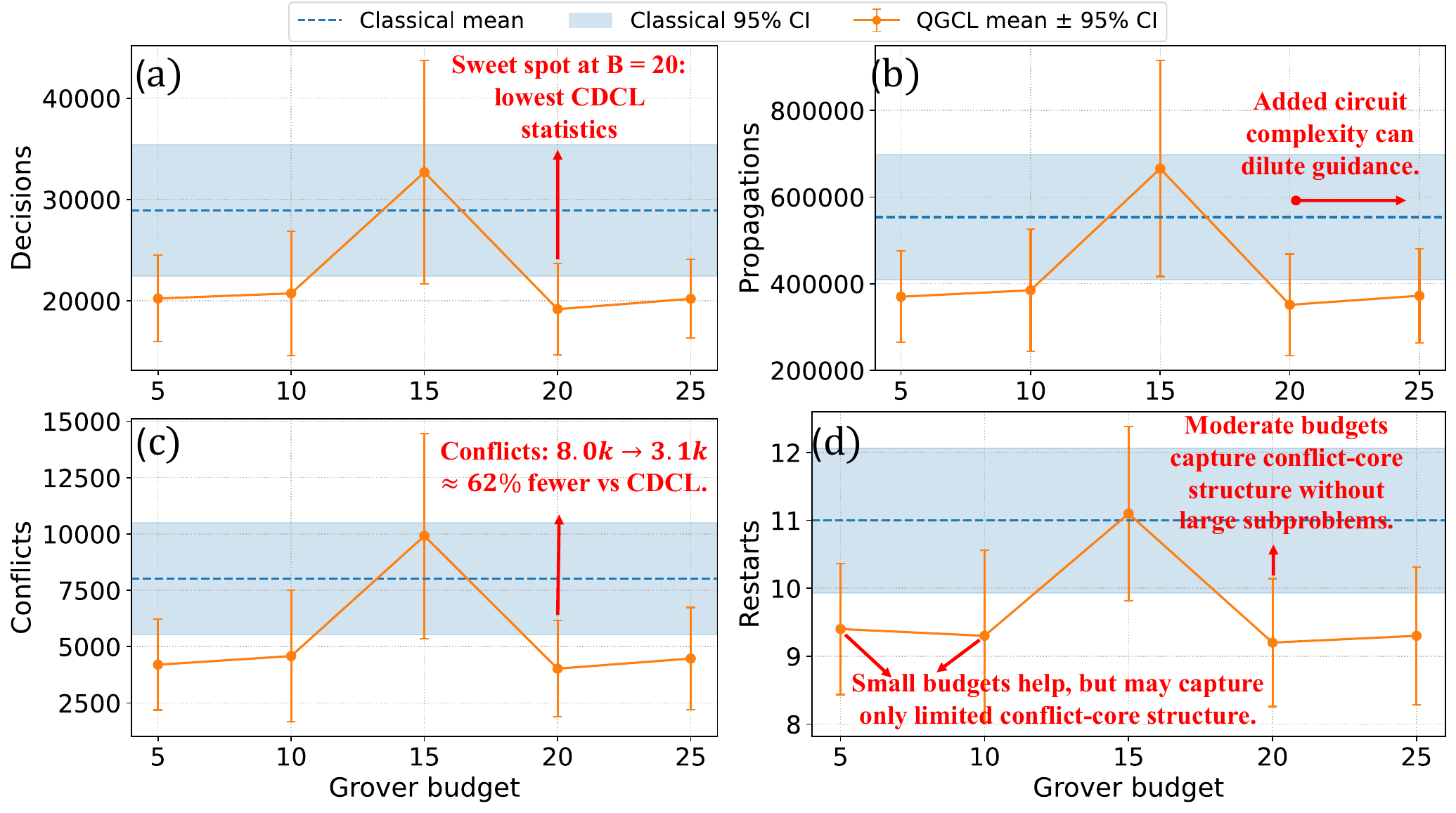}
    \caption{QGCL performance when varying the Grover budget, defined as the maximum size of the extracted subformula (number of distinct variables plus clauses), on the same fixed instance used in the call-count sweep. Panels show (a) decisions, (b) propagations, (c) conflicts, and (d) restarts. The classical CDCL baseline mean and 95\% confidence interval are plotted in blue; QGCL (hybrid) means and 95\% confidence intervals appear in orange.}
    \label{fig:GroverBudget}
    \vspace{-15pt}
\end{figure}

In the third experiment, we fix the number of Grover calls and sweep the Grover budget $B$, defined as the maximum extracted subformula size measured by the number of distinct variables plus the number of clauses. In our logical oracle model, this budget roughly tracks the circuit width because $W_{\mathrm{logical}}=B+1$ before backend-specific decomposition. Figure~\ref{fig:GroverBudget} shows the trends for a fixed instance $T=2$, including both the classical CDCL baseline and the hybrid QGCL solver, with $95\%$ CIs. Table~\ref{tab:t2_sweeps} reports the corresponding simulator budget-sweep values.

The updated budget sweep shows that QGCL is sensitive to the extracted-subformula budget. The best setting is $B=20$, which gives the lowest mean restarts, conflicts, decisions, propagations, and wall time in the simulator budget block ($3.96$~s versus $5.88$~s for CDCL). Smaller budgets ($B=5$ and $10$) provide useful but weaker guidance, while larger budgets do not automatically improve performance. This supports the view that QGCL benefits from a tuned, moderate subformula size: large enough to capture conflict-core structure, but not so large that oracle construction, BBHT randomness, or simulation cost dominate.

The wall-time improvement is especially notable because the present implementation still simulates Grover circuits. BBHT is used because the number of satisfying assignments in each extracted CNF is unknown, and it may require several randomized iteration-count guesses before a useful candidate is found. The results indicate that, in the noiseless simulator, the search reduction can outweigh this overhead. This strengthens the evidence for QGCL as a useful guidance mechanism, while leaving open the separate question of how much overhead remains on noisy hardware.

The FakeBackend rows in Table~\ref{tab:t2_sweeps} provide an initial noisy-backend sanity check. With $B=10$, QGCL reduces conflicts from $8{,}016.20$ to $4{,}459.80$, decisions from $28{,}905.30$ to $21{,}062.60$, and propagations from $553{,}733.80$ to $372{,}370.20$, but increases wall time from $6.11$ s to $80.81$ s. This overhead reflects both the cost of noisy simulation and the fact that backend noise can blur Grover-amplified peaks that BBHT relies on. In such cases, noise may delay or prevent the procedure from quickly identifying a satisfying assignment, causing extra attempts.

The qualitative histograms in Figure~\ref{fig:GroverSAT}(c)-(d) are consistent with this table-level behavior. The noisy backend runs show that noise can introduce spurious high-probability candidates, but the QGCL feedback path is deliberately conservative: only top-ranked candidates are considered, they are scored by the classical subformula checker, and they affect only activities and polarity preferences. This gives the method limited noise tolerance as a guidance mechanism, while preserving the earlier limitation that noisy simulation can increase wall time and reduce the reliability of each Grover call.

\begin{figure}
    \centering
    \includegraphics[width=1\linewidth]{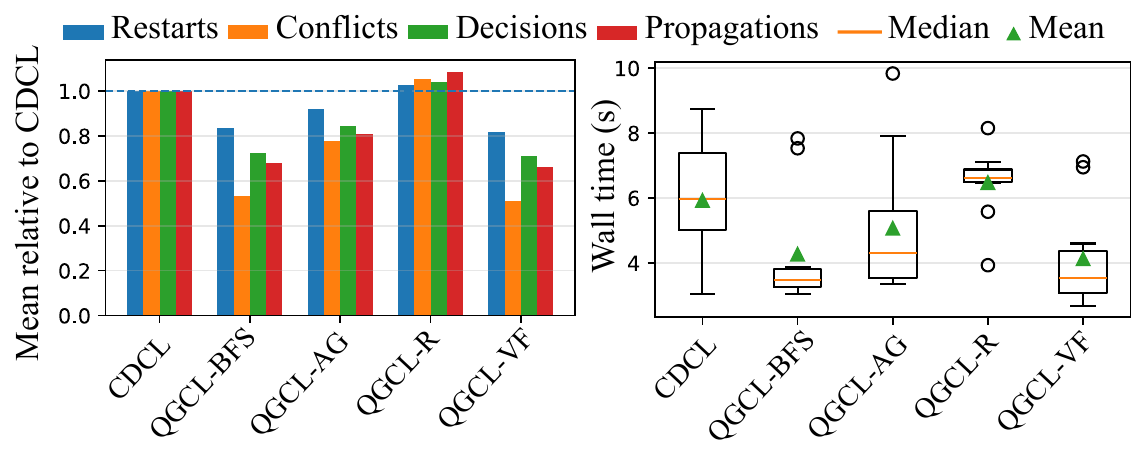}
\caption{Extraction strategies in QGCL on the \(T=2\) Power SCA CNF instance. Structured localization, especially activity-BFS and variable-frontier (VF), reduces search effort and wall time more effectively than random extraction.}
\label{fig:extractors}
\vspace{-15pt}
\end{figure}

%%%%%%%%%%%%%%%%%%%%%%%%%%%%%%%%%%%%%%%%%%%%%%%%%%%%%%%%%%%%%%%%%%%%%%%%%%%%%%%%%%%%%%%%%%%%%%%%%%%%%%
Additionally, Figure~\ref{fig:extractors} shows that the extraction strategy strongly affects QGCL performance on the \(T=2\) Power SCA CNF instance. Activity-BFS expands from a high-conflict clause through variable-sharing clauses, activity-greedy (AG) selects the highest-activity clauses, random (R) samples clauses uniformly, and variable-frontier (VF) gathers clauses around high-activity variables. The structured methods outperform random extraction, with activity-BFS and variable-frontier giving the strongest gains. while random extraction can increase search effort and wall time because it is less likely to isolate a coherent conflict-relevant subproblem.

The performance gains are primarily attributable to the coupling of structural localization and Grover search. QGCL first localizes subformulas that are both complex and structurally central, thereby restricting quantum search to regions where it is most likely to yield an advantage. As Figure~\ref{Complexity} indicates, Grover search becomes effective only when the localized subformula attains sufficient size and structural significance, making targeted quantum acceleration preferable to a generic classical SAT-based treatment of the subproblem.
%%%%%%%%%%%%%%%%%%%%%%%%%%%%%%%%%%%%%%%%%%%%%%%%%%%%%%%%%%%%%%%%%%%%%%%%%%%%%%%%%%%%%%%%%%%%%%%%%%%%%%

Therefore, the QGCL framework is NISQ-compatible in architecture because it restricts the quantum workload to small, dynamically selected subformulas and keeps final correctness classical. It is not yet NISQ-efficient in wall time. Addressing this limitation will require noise-aware oracle synthesis, transpilation-aware budgets, adaptive BBHT stopping rules, fewer shots when the histogram is already decisive, and error-mitigation or calibration strategies tailored to the chosen backend.

\vspace{-10pt}
\section{Conclusion}

This work investigated how a quantum routine can be integrated into a state-of-the-art SAT solver for SAT-assisted AES-128 power side-channel cryptanalysis. We introduced QGCL, a hybrid solver that embeds Grover search calls inside a CDCL engine and applies them only to small CNF subformulas extracted from conflict cores. The quantum layer acts purely as a guide that proposes locally consistent assignments and refines branching heuristics, while the final logical decision remains classical.

On our AES-oriented power-SCA benchmark family with leakage-derived hint configurations, QGCL reduces solver-internal statistics on several structurally hard instances, including benchmarks with up to $39{,}389$ variables and $137{,}712$ clauses. Conflict counts improve by up to $86\%$ relative to the CDCL baseline, with similar trends for other statistics. Parameter sweeps over the maximum number of Grover calls and the subproblem budget reveal a useful operating window: moderate budgets and enough calls produce strong guidance, while larger settings show diminishing returns.

The wall-time results are encouraging in the ideal simulator: QGCL has lower mean wall time than CDCL across the updated depth sweep and in the best $T=2$ sweep settings. Because the implementation still simulates Grover circuits and uses BBHT, we treat these timings as prototype evidence rather than a hardware claim. The noisy backend experiment further shows that noise can preserve search-statistic reductions while substantially increasing runtime.
These results demonstrate that targeted quantum guidance can reduce CDCL search effort and, in the updated ideal-simulator measurements, can improve mean wall time. The noisy backend results further clarify the path forward, highlighting noisy-hardware performance as a promising next milestone rather than a limitation of the approach.

Future work should extend the evaluation to richer stream-cipher and block-cipher CNFs, more realistic leakage models, and fully learned guidance policies. It should also compare QGCL with other quantum paradigms, such as QAOA or quantum annealing, under the same hardware budget. Finally, deploying QGCL in noisy environments will require noise-aware oracle design, backend-specific transpilation, adaptive BBHT policies, and error mitigation. Overall, QGCL provides a concrete, quantitatively supported blueprint for using quantum resources where they matter most in SAT-based cryptanalysis: not to replace CDCL, but to steer it faster toward more productive conflict cores.

\vspace{-5pt}
\section*{Acknowledgments}
\vspace{-5pt}
This work was supported in part by the NYUAD Center for Cyber Security (CCS), funded by Tamkeen under the NYUAD Research Institute Award G1104, and by the NYUAD Center for Quantum and Topological Systems (CQTS), funded by Tamkeen under the NYUAD Research Institute grant CG008.

\bibliographystyle{IEEEtran}
\bibliography{main}
\end{document}